\documentclass[11pt,aps,nofootinbib,preprintnumbers,citeautoscript]{revtex4}
\usepackage{amsmath,amssymb}
\usepackage{enumerate}
\usepackage{graphicx}
\usepackage{xcolor}
\usepackage{enumerate}
\usepackage{gensymb}
\usepackage[colorlinks=true,citecolor=red,linkcolor=,linkbordercolor=cyan]{hyperref}

\usepackage{float}
\date{\today}
\begin{document}
\title{Entanglement sudden death and birth effects in two qubits maximally entangled mixed states under quantum channels}

\author{Kapil K. Sharma$^{\dagger}$ and Vladimir P. Gerdt$^{\pm}$\\
\textit{Laboratory of Information Technologies, \\ Joint Institute for Nuclear Research, \\ 6 Joliot-Curie St, Dubna, 141980,\\
Moscow Oblast, Russian Federation} \\
E-mail: iitbkapil@gmail.com$^{\dagger}$, gerdt@jinr.ru$^{\pm}$}

\begin{abstract}
In the present article, the robustness of entanglement in two qubits maximally entangled mixed states (MEME) have been studied under quantum decoherence channels. Here we consider bit flip, phase flip, bit-phase-flip, amplitude damping, phase damping and depolarization channels. To quantify the entanglement, the concurrence has been used as an entanglement measure. During this study interesting results have been found for sudden death and birth of entanglement under bit flip and bit-phase-flip channels. While amplitude damping channel produces entanglement sudden death and does not allow re-birth of entanglement. On the other hand, two qubits MEMS exhibit the robust character against the phase flip, phase damping and depolarization channels. The elegant behavior of all the quantum channels have been investigated with varying parameter of quantum state MEMS in different cases.
 
\end{abstract}
\maketitle
\section{Introduction}
Quantum systems may carry quantum correlations in different physical conditions. These quantum correlations are the key stone to develop quantum applications in may domains like quantum cryptography, quantum game theory, quantum imaging, quantum algorithms, and quantum machine learning, quantum metrology etc \cite{qc1,qgame1,qi1,qalgo1,qaml1,qmet1}. Till today quantum community is well familiar with two kinds of quantum correlations such as entanglement and quantum discord\cite{ent1,qd1}. The first one (entanglement) is highly investigated with theoretical as well as experimental studies, the second one is  measurement based quantum correlation and deep investigations for quantum discord are on the way. Here we mention that manipulating the entanglement in varieties of quantum states is always the subject of study for quantum information community. Obviously sustainability of entanglement in quantum systems for a long time is the experimental requirement to develop quantum applications. Quantum information community is always interested to search such quantum systems which persist entanglement for a long time. For example, transmitting the quantum information from source to destination require a long connection of optical fibers in which entangled photos are launched\cite{qint1,qint2,qint3}. If entanglement in photons do not sustain over the long distance than quantum communication will break. Hence generally for long distance quantum communication the concept of quantum repeaters\cite{qmrep1}  and quantum memory\cite{qmm} has been developed, which may be assumed as analogues to the amplifiers over the classical channels. Quantum repeaters help to protect the decay of entanglement and propagate it over the quantum channel. Here it is important to protect the vanishing of entanglement over a long distance. Quantum communication is not only the example, but sustainability of entanglement is also required in condensed matter physics along with the thermal environment as well. Most of the time decoherence is the main factor which adversely influence the entanglement, because quantum systems are too evasive and decoherence prone everytime. If entanglement dies in quantum systems by any means,  than it becomes very difficult to use such systems for application. The vanishing of entanglement for a finite time is called entanglement sudden death (ESD)\cite{esd1,esd2}. This phenomenon is first time observed by Yu and Eberly dealing with the time evolution of two qubits in cavity systems. After that, quantum information literature cover a lot of investigations of this phenomenon under various conditions which are observed since last many years\cite{k1,k2,k3,k4,k5,k6}.  In this line of research, it is essential to avoid or delay the ESD to develop better quantum applications. There are few schemes  to protect the entanglement from decoherence such as week measurement in association with quantum measurement reversal\cite{mw1,mw2,mw3,mw4,mw5,mw6,mw7,mw8,mw9}, feed back control\cite{feed1,feed2}, unitary operations\cite{un1}, delayed choice decoherence suppression\cite{dch1}, dynamical decoupling\cite{dcop1,dcop2} and decoherence free subspace\cite{dfs1,dfs2}, quantum zeno effect\cite{qz1,qz2} . In general, there is no generalized method to protect ESD in quantum systems. But it is always interesting to investigate the situations of ESD and observe it in varieties of quantum states, so that it may be useful for experimental quantum information community. In parallel, it will not be worth to mention that, many time in the absence of entanglement the quantum discord exists in the physics systems, which may be utilized to develop the quantum applications. But experimental studies on practical execution of quantum discord for varieties of applications are limited and investigations are on the way. Here in the present article we invistigate the dynamical behavior of two qubits maximally mixed entangled state (MEMS)\cite{mun1} under quantum decoherence channels. Following the literature on two qubits maximally mixed entangled states (MEMS), we have found that state have more amount of entanglement than two qubits Werner state\cite{wer1}. Hence  MEMS is important for quantum information community. In the present study we use concurrence as an entanglement measure and  
well known quantum decoherence channels such as bit flip, phase flip, bit-phase-flip, amplitude damping, phase damping, and depolarization\cite{channels}. 

\section{Concurrence, MEMS and Werner state.}
In this section deal with two subsections. In first subsection we present the mathematical method to calculate the entanglement by using a measure called concurrence. In second subsection we present the mathematical structure of density matrix of two quibts MEMS and its concurrence in comparison to Werner state.
\subsection{Concurrence}\label{con}
The concurrence $C$ in a density matrix 
$\rho$ is given by 
\begin{equation}
C(\rho)=max\{0,\lambda_{1}-\lambda_{2}-\lambda_{3}-\lambda_{4}\},
\end{equation}
where $\lambda_{i}$, $i=1,2,3,4$, are the square roots of the eigenvalues in decreasing order of $\rho \rho^{f}$, $\rho^{f}$ is the spin flip density matrix given as \\
\begin{equation}
\rho^{f}=(\sigma^{y}\otimes \sigma^{y}) \rho^{*}( \sigma^{y}\otimes \sigma^{y}). 
\end{equation}
Here $\rho^{*}$ is the complex conjugate of the density matrix $\rho$ and $\sigma^{y}$ is the Pauli Y matrix. The eigenvalue spectrum of the matrix $\rho \rho^{f}$ is used to calculate the concurrence throughout the paper.
\subsection{Concurrence in MEMS and Werner states: A comparison}
Two qubit Werner states can be written as 
\begin{equation}
\rho_{w}=\gamma |\psi^{-}\rangle \langle \psi^{-}|+(1-\gamma)\frac{I}{4}. \label{eq:1}
\end{equation}
where $|\psi^{-}\rangle$ is the singlet state and $I$ is the $4 \times 4$ identity matrix. This state is entangled with $\gamma>1/3$ and carry the concurrence as $C(\gamma)=(3\gamma -1)/2$.  On the other hand the two qubit MEMS have been investigated by Munro et al., by introducing a function of the parameter $g(\gamma)$ along the diagonal of the density matrix. He found the cases with $g(\gamma)$, which makes the state more entangled than Werner state. The density metrics of MEMS is given below,
\begin{center}
\begin{equation}
\rho_{MEMS}= \left[ \begin{array}{cccc}
          g(\gamma) &0 &0 &\gamma/2   \\
           0 &1-2 g(\gamma)&0&0     \\
           0&0&0&0     \\
           \gamma/2&0&0&g(\gamma)     \\
       \end{array}
      \right ]; \quad g(\gamma)=\begin{cases}
        1/3,              & 0 \leq\gamma < 2/3\\
        \gamma /2,& 2/3 \leq \gamma\leq 1
\end{cases} \label{cs}
\end{equation} 
\end{center}
These states carry the concurrence $C(\gamma)=\gamma$. We compare the concurrences in both the states as depicted in Fig. 1. 
\begin{figure}
\includegraphics[scale=1.2]{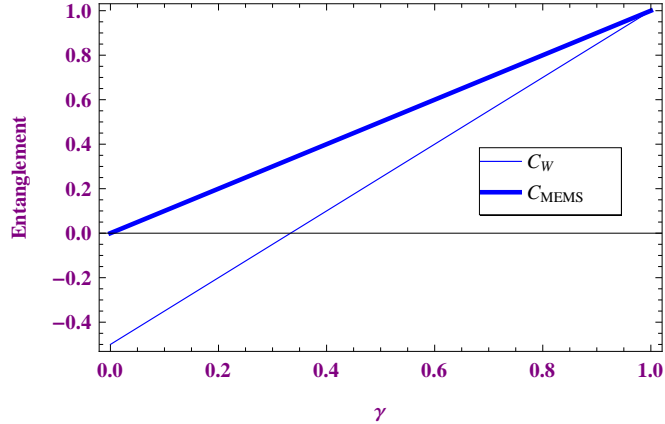}
\caption{A comparison of concurrence in MEMS and Werner state: $C_{W}$(Concurrence in Werner state), $C_{MEMS}$(Concurrence in MEMS ) }
\end{figure}
In this figure $C_{W}$ represent the concurrence in Werner states and $C_{MEMS}$ represents the concurrence  in MEMS. The concurrence achieves negative values for Werner states with $\gamma <1/3$, so it is disentangle within the range $0 \leq \gamma <1/3$, while on the other the hand MEMS carry more concurrence than Werner states and hence more entangled than Werner states. Here we elucidate that, we focus on the decoherence study with the two cases of MEMS given in \ref{cs}. So, in this direction our study bifurcate in two cases as impact of quantum decoherence channels on MEMS with $(0\leq \gamma <2/3)$ and with $(2/3\leq \gamma <1)$. The former is treated as case 1 and later one is considered as case 2.  

\section{Decoherence prone density matrix and Quantum channels.}
Under this section we develop the mathematical equation of decoherence prone density matrix with different quantum decoherence channels. Here we consider that, each qubit passes through dcoherence independently. So following this, the mathematical equation of decoherence prone density matrix can be written as,
\begin{equation}
\rho^{dp}=\sum_{i=1}^{n}\sum_{j=1}^{n}E_{i}\otimes E_{j}[\rho_{MEMS}]E_{i}^{\dagger}\otimes E_{j}^{\dagger}. \label{dec}
\end{equation}
\begin{figure*}
\includegraphics[scale=1.0]{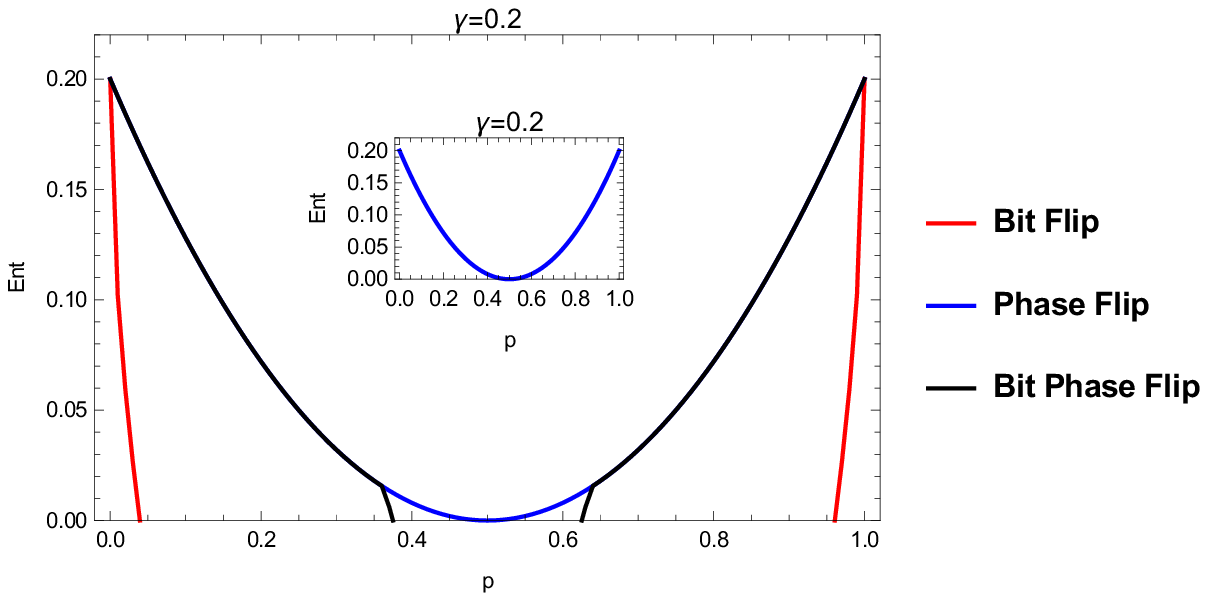}
\caption{Case 1:} \label{rc1}
\end{figure*}

\begin{figure*}
\includegraphics[scale=1.0]{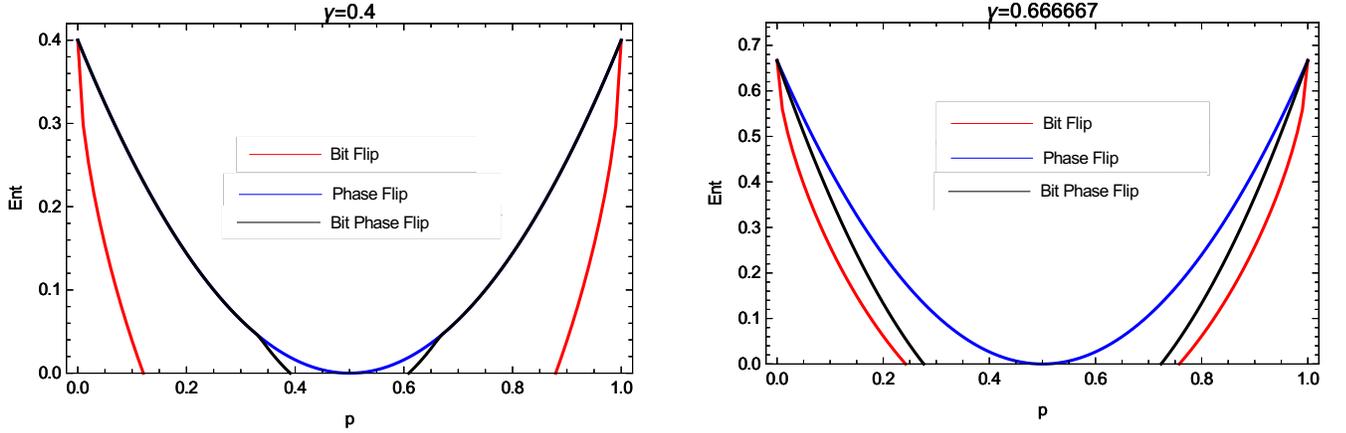}
\caption{Case 1: Entanglement plot for bit flip, phase flip, bit-phase-flip channels with $(0\leq \gamma< 2/3)$} \label{rc2}
\end{figure*}
with the conditions,
\begin{eqnarray}
\sum_{i}^{n}E_{i}E_{i}^{\dagger}=I.\\
\sum_{j}^{n}E_{i}E_{j}^{\dagger}=I.
\end{eqnarray}
Where $\rho_{MEMS}$ is the initial density matrix of the quantum state and $n$ is the number of Kraus operators corresponding to a particular decoherence channel. The Kraus operators of bit flip, phase flip, bit-phase-flip, amplitude damping, phase damping and depolarization channels are provided in Appendix \ref{aa}.
\section{Case 1: $(0 \leq \gamma<2/3)$}
In this section we consider the case 1, as $(0\leq \gamma<2/3)$, under quantum decoherence channels. We begin our study with different quantum channels declared as bit flip, phase flip, bit-phase-flip, amplitude damping, phase damping, depolarization channels. The study is carried out in subsequent subsections.
\subsection{Bit flip, Phase flip and Bit-phase-flip channel}
We plug the Kraus operators of bit flip, phase flip and bit-phase-flip channels in Eq.\ref{dec} and find out the eigenvalues spectrum of the  density matrix $\rho^{dp}(\rho^{dp})^{f}$ by following the procedure given in subsection \ref{con}. We present the corresponding eigenvalues spectrum of the density matrix in subsections \ref{bfc},\ref{pfc} and \ref{bpfc}. These eigenvalues spectrum appear at the end of this subsection. Looking into the eigenvalue spectrum, we observe that concurrence is the function of two variables $p$ and $\gamma$. Where $p$ is the decoherence parameter and $\gamma$ is the parameter of quantum state MEMS, the different values of the parameter $\gamma$ produce different two qubits MEMS. Further we simulate the function of concurrence and do analytical study of the behavior of above mention channels. The combine plots of entanglement has been shown in Fig.\ref{rc1} and \ref{rc2} with varying values of the parameter $\gamma$. Observing the plot in Fig.\ref{rc1} with the parameter value $\gamma=0.2$, we find the maximal entanglement achieved in the state is $0.20$ for all the quantum channels. Here we recall that, the maximum entanglement carried out in MEMS is $C(\gamma)=\gamma$, hence  our simulated result for maximum entanglement for $(\gamma=0.2)$ is verified as correct, which also can be checked with subsequent cases as true. With $(\gamma=0.2)$, we observe that bit flip and bit-phase-flip channels produce ESD at $(p=0.049)$ and $(p=0.375)$ respectively. But entanglement takes re-birth in the state again with $(p=0.952)$ and $(p=0.625)$ for bit flip and bit-phase-flip channels respectively. Here entanglement sudden death and birth (ESDB) effect have been observed. The zone of ESDB for bit flip channel is wider than bit-phase-flip channel. Giving the insights to the Fig. \ref{rc2}, as the value of the parameter $\gamma$ increases with $(\gamma=0.4)$, the amount of the entanglement in the state increased to $0.4$ and simultaneously, the ESDB zone is also decreased for bit flip and bit-phase-flip channels. On the other hand with $(\gamma \approx 0.66)$, again the entanglement in the state rises to $0.666667$ and ESDB zone squeezed for bit flip channel but it is increased for bit-phase-flip channel. Here we find the results that, phase flip channel do not disturb the quantum states with $(0 \leq \gamma< 2/3)$, but bit flip channel and phase flip channel has typical influence on the quantum states and produce the squeezing ESDB zones. The length of ESDB zones are given in Eq.\ref{esz}. 
\begin{eqnarray}
\text{Bit Flip}\rightarrow \begin{cases}
    \gamma=0.2 ,& (0.049-0.952)^{\star} \\
    \gamma=0.4, & (0.125-0.8525)^{\star} \\
\gamma=0.66, & (0.225-0.75)^{\star} 
\end{cases}; \quad \text{Bit-phase-flip}\rightarrow \begin{cases}
    \gamma=0.2 ,& (0.375-0.625)^{\star} \\
    \gamma=0.4, & (0.4-0.6)^{\star} \\
\gamma=0.66, & (0.3-0.725)^{\star} 
\end{cases}\label{esz}  
\end{eqnarray}
Where $(.)^{\star}$ represents the ESDB zone length.
\begin{figure}
\includegraphics[scale=1.0]{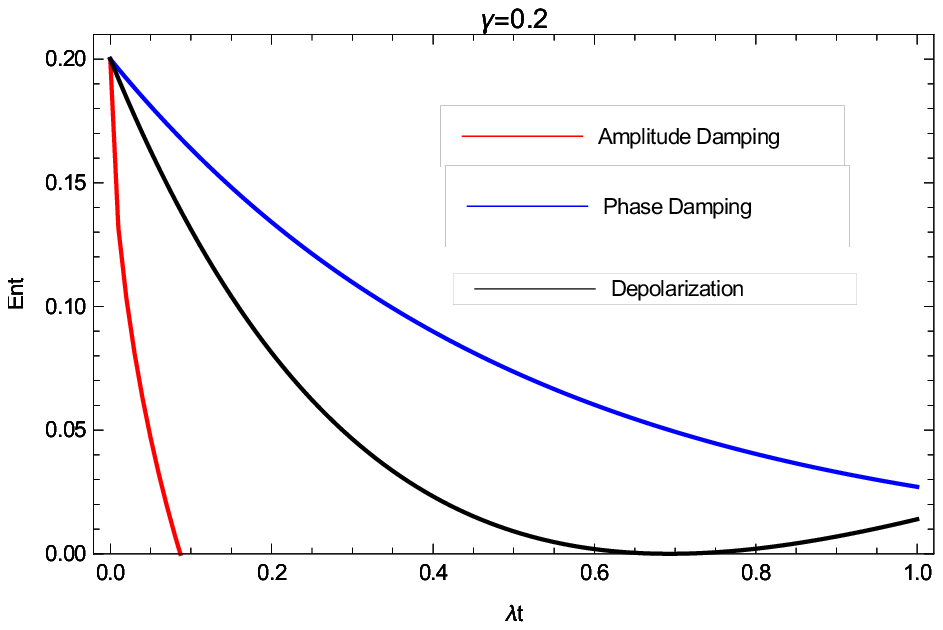}
\caption{Case 1: Entanglement plot for amplitude damping, phase damping and depolarization channels with $(0\leq \gamma < 2/3)$.}\label{a1}
\end{figure}

\begin{figure}
\includegraphics[scale=0.8]{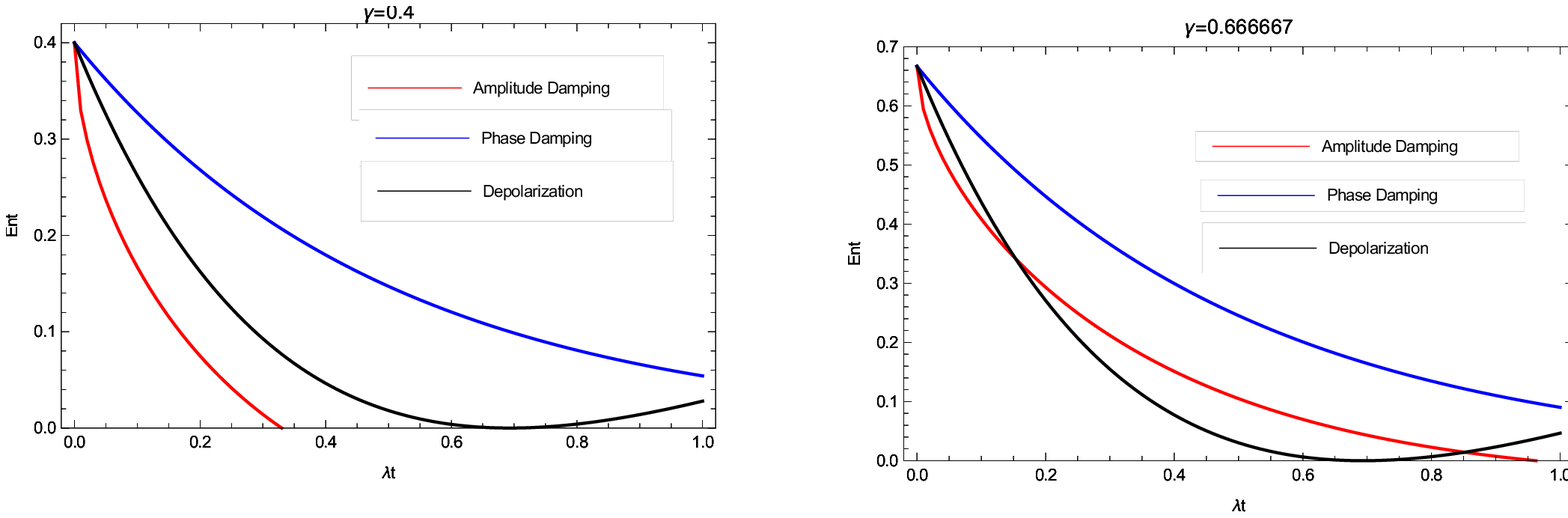}
\caption{Case 1:Entanglement plot for amplitude damping, phase damping and depolarization channels with $(0\leq \gamma < 2/3)$.}\label{a2}
\end{figure}
\subsubsection{Eigenvalue Spectrum of $\rho^{dp}(\rho^{dp})^{f}$: Bit flip channel}\label{bfc}
\begin{equation}
\{\lambda_{1},\lambda_{2} \}=\{\frac{1}{36} (3 \gamma +2 (3 \gamma -1) (p-1) p-2)^2,\frac{1}{36} (3 \gamma +2 (3 \gamma +1) (p-1) p+2)^2\}
\end{equation}

\begin{equation}
\lambda_{3}=\frac{1}{9} \left(p \left(p
\left(p^2 +9 \gamma^2 (p-1)^2-2 p-1\right)+2\right)-6 \sqrt{\gamma^2 (p-2) (p-1)^3 p^3 (p+1)}\right),
\end{equation}

\begin{equation}
\lambda_{4}=\frac{1}{9}  \left(6 \sqrt{\gamma^2 (p-2)(p-1)^3 p^3 (p+1)}+p \left(p \left(p^2+9 \gamma^2 (p-1)^2-2 p-1\right)+2\right)\right)\}
\end{equation}

\subsubsection{Eigenvalue Spectrum of $\rho^{dp}(\rho^{dp})^{f}$: Phase flip channel}\label{pfc}
\begin{equation}
\{\lambda_{1},\lambda_{2},\lambda_{3},\lambda_{4}\}=\{0,0,\frac{1}{36} \left(2-3 \gamma (1-2 p)^2\right)^2,\frac{1}{36} \left(3 \gamma (1-2 p)^2+2\right)^2\}
\end{equation}

\subsubsection{Eigenvalue Spectrum of $\rho^{dp}(\rho^{dp})^{f}$: Bit-Phase-Flip channel}\label{bpfc}
\begin{equation}
\{\lambda_{1},\lambda_{2}\}=\{\frac{1}{36} (3 \gamma +6 (\gamma -1) (p-1) p-2)^2,\frac{1}{36} (3 \gamma +6 (\gamma +1) (p-1) p+2)^2\}
\end{equation}

\begin{equation}
\lambda_{3}=\frac{1}{9} \left((p-1) p \left(9 \left(\gamma^2+1\right) (p-1) p+2\right)-6 \sqrt{\gamma^2 (p-1)^3 p^3 (9 (p-1) p+2)}\right).
\end{equation}

\begin{equation}
\lambda_{4}=\frac{1}{9} \left(6 \sqrt{\gamma^2 (p-1)^3 p^3 (9 (p-1) p+2)}+(p-1) p \left(9 \left(\gamma^2+1\right) (p-1) p+2\right)\right)\}.
\end{equation}

\subsection{Amplitude damping, Phase damping and Depolarization channels }
Under this subsection we discuss the combine study of amplitude damping, phase damping and depolarization channels with $(0\leq\gamma<2/3)$. By using the similar procedure mentioned in the previous section we first need to find the eigenvalues spectrum of the decoherence prone density matrix $\rho^{dp}(\rho^{dp})^{f}$. The eigenvalues spectrum for all the three channels are given in subsections \ref{adc},\ref{pdc} and \ref{dc}, which is further used to calculate the concurrence in the state. These eigenvalues spectrum appear at the end of this subsection. Here eigenvalues spectrum involve  the parameters $(\lambda t)$ and $(\gamma)$,  where $(\lambda t)$ is the damping rate of the channels.
The graphical results are presented in Fig.\ref{a1} and \ref{a2}. Here we would like to mention that all the three channels have totally different behavior than bit flip, phase flip and bit-phase-flip channels as the damping rate $(\lambda t)$ increases. Observing the Fig.\ref{a1} with the parameter value $\gamma=0.2$, we investigate the highest entanglement is achieved is $(0.2)$ for all the quantum channels, which again full fill the condition of concurrence $\large(C(\gamma)=\gamma\large)$ and verify that the simulated results are correct. As the parameter $(\lambda t)$ increases the amplitude damping channel loose the entanglement in two qubits MEMS and produce ESD, the entanglement do not appear with $(\lambda t)>0.99$. On the other hand phase damping and depolarization channels do not loose the entanglement from two qubits MEMS, but slowly decaying it as the value of the parameter $(\lambda t)$ increasing. Depolarization channel makes the entanglement zero for finite range of $(\lambda t)$, but do not produce the ESD. Further when the value of $(\gamma)$ increases, the amount of entanglement in two qubits MEMS increases. With the increasing value of $(\gamma)$, the amplitude damping channel has the tendency to regain the entanglement with the increasing value of damping parameter $(\lambda t)$. Further it is completely recovered when $(\gamma)$ approaches to unity. The results of ESD zone length for amplitude damping channels are given in Eq.\ref{am1}. Comparing the influence of all the three channels, it has been found that the two qubits MEMS is sensitive with amplitude damping and depolarization channel, while it is robust against the phase damping channel.
\begin{equation}
\text{Amplitude damping}\rightarrow \begin{cases}
    \gamma=0.2,& (\lambda t=0.99)^{\dagger} \\
    \gamma=0.4, & (\lambda t=0.325)^{\dagger} \\\end{cases} \label{am1}
\end{equation}
Where $(.)^{\dagger}$ represent the starting point of ESD zone.
\subsubsection{Eigenvalue Spectrum of $\rho^{dp}(\rho^{dp})^{f}$: Amplitude damping channel}\label{adc}
\begin{equation}
\{\lambda_{1},\lambda_{2}\}=\{\frac{1}{9} e^{-4 \text{$\lambda $t}} \left(e^{\text{$\lambda $t}}-1\right) \left(2 e^{\text{$\lambda $t}}-1\right), \frac{1}{9} e^{-4
   \text{$\lambda $t}} \left(e^{\text{$\lambda $t}}-1\right) \left(2 e^{\text{$\lambda $t}}-1\right)\}
\end{equation}

\begin{equation}
\lambda_{3}=\frac{1}{36} e^{-12 \text{$\lambda $t}}
   \left(e^{8 \text{$\lambda $t}} \left(3 e^{\text{$\lambda $t}} \left(\left(3 \gamma ^2+4\right) e^{\text{$\lambda $t}}-4\right)+4\right)-12 \sqrt{\gamma ^2 e^{19 \text{$\lambda $t}} (2 \sinh (\text{$\lambda $t})+4 \cosh (\text{$\lambda $t})-3)}\right)
\end{equation}

\begin{equation}
\lambda_{4}=\frac{1}{36} e^{-12 \text{$\lambda
   $t}} \left(e^{8 \text{$\lambda $t}} \left(3 e^{\text{$\lambda $t}} \left(\left(3 \gamma ^2+4\right) e^{\text{$\lambda $t}}-4\right)+4\right)+12
   \sqrt{\gamma ^2 e^{19 \text{$\lambda $t}} (2 \sinh (\text{$\lambda $t})+4 \cosh (\text{$\lambda $t})-3)}\right)\}
\end{equation}

\subsubsection{Eigenvalue Spectrum of $\rho^{dp}(\rho^{dp})^{f}$: Phase damping channel}\label{pdc}
\begin{equation}
\{\lambda_{1},\lambda_{2},\lambda_{3},\lambda_{4}\}=\{0,0,\frac{1}{36} e^{-4 \text{$\lambda $t}} \left(2 e^{2 \text{$\lambda $t}}-3 \gamma \right)^2,\frac{1}{36} e^{-4 \text{$\lambda $t}} \left(3
   \gamma +2 e^{2 \text{$\lambda $t}}\right)^2\}
\end{equation}

\subsubsection{Eigenvalue Spectrum of $\rho^{dp}(\rho^{dp})^{f}$: Depolarization channel}\label{dc}
\begin{equation}
\{\lambda_{1},\lambda_{2}\}= \{\frac{4}{81} \gamma ^2 e^{-4 \text{$\lambda $t}} \left(-5 e^{\text{$\lambda $t}}+e^{2 \text{$\lambda $t}}+4\right)^2,\frac{4}{81} \gamma ^2
   e^{-4 \text{$\lambda $t}} \left(-5 e^{\text{$\lambda $t}}+e^{2 \text{$\lambda $t}}+4\right)^2\}
\end{equation}

\begin{equation}
\lambda_{3}=\frac{e^{-2 \text{$\lambda $t}} ((45 \gamma -6)
   \sinh (\text{$\lambda $t})+(10-75 \gamma ) \cosh (\text{$\lambda $t})+48 \gamma +8)^2}{2916}.
\end{equation}

\begin{equation}
\lambda_{4}=\frac{e^{-4 \text{$\lambda $t}}
   \left(e^{\text{$\lambda $t}} \left((15 \gamma +2) e^{\text{$\lambda $t}}-48 \gamma +8\right)+60 \gamma +8\right)^2}{2916}
\end{equation}

\begin{figure*}
\includegraphics[scale=1.0]{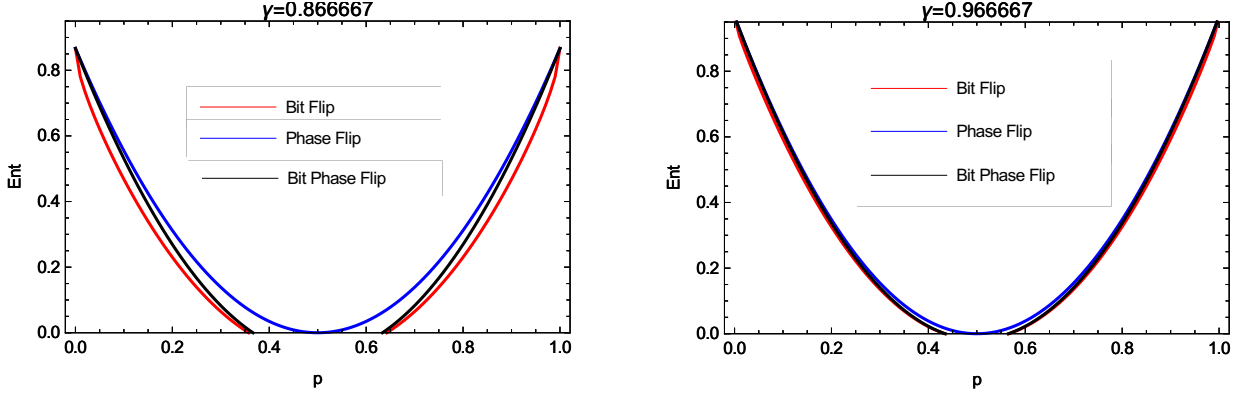}
\caption{Case 2: Entanglement plot for bit flip, phase flip, bit-phase-flip channels with $(2/3\leq \gamma \leq 1)$} \label{rc3}
\end{figure*}

\section{Case 2: $(2/3\leq \gamma \leq 1)$}
In this section we present the study for case 2 ie. $(2/3\leq \gamma \leq 1)$. First we discuss the behavior of bit flip, phase flip, bit-phase-flip channels and later amplitude damping, phase damping and depolarization channels.
\subsection{Bit flip, Phase flip and Bit-phase-flip channel}
In this section we consider the case $(2/3\leq \gamma \leq 1)$. The graphical results for this case have been shown in Fig.\ref{rc3}. We have found as the value of the parameter $\gamma$ increases beyond $2/3$, the ESDB zone keeps on squeezing for both bit flip and bit-phase-flip channels. For $\gamma=0.888887$ the length of ESDB zone is observed as $(0.375-0.625)$. It is important to note that bit flip and bit-phase-flip channels produce the same length of ESD zone. As the value of $\gamma$ approach to unity, the ESD zone completely vanish, but it still exists with $\gamma=0.999997$. The length of ESDB zone for different values of $\gamma$ is given in Eq.\ref{cs2}
\begin{equation}
\text{Bit flip, bit-phase-flip}\rightarrow \begin{cases}
    \gamma=0.866 ,& (0.375-0.625)^{\star} \\
    \gamma=0.9666, & (0.445-0.525)^{\star} 
\end{cases}\label{cs2} 
\end{equation}
So the MEMS corresponding to $\gamma=1$, does not suffer from ESD and exhibit the robust character. In this case also, the phase flip channel do not disturb the two qubits MEMS.
\begin{figure}
\includegraphics[scale=1.0]{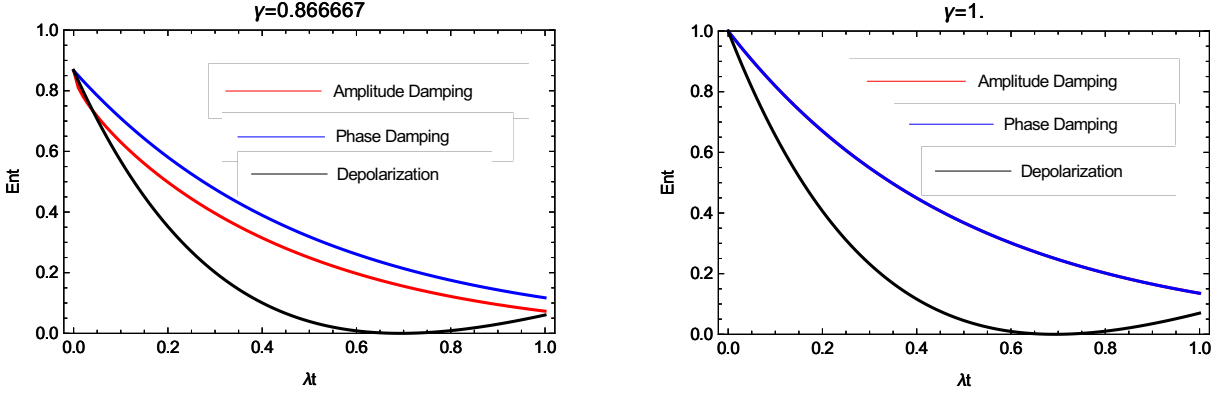}
\caption{Case 2: Entanglement plot of amplitude damping, phase damping and depolarization channels with $(2/3\leq \gamma \leq 1)$}\label{a3}
\end{figure}
\subsection{Amplitude damping, Phase damping and Depolarization channel}
Under this subsection we discuss influence of amplitude damping, phase damping and depolarization channels with the case $(2/3\leq\gamma\leq1)$. The eigenvalue spectrum is provided in subsections \ref{adc},\ref{pdc} and \ref{dc} and graphical results are shown in Fig.\ref{a3}. Observing the graphical results, we find as the value of parameter $(\gamma)$ adopt the range $\gamma>2/3$, the amount of entanglement in the state increases following the property $\large(C(\gamma)=\gamma \large)$. On the other hand 
as the decoherece rate $(\lambda t)$ increases the behavior of amplitude damping channel changes, it recover the entanglement by avoiding ESD and finally have the same behavior as phase damping channel with $(\gamma=1)$. Here we conclude that, non of the three channels produce ESD in two qubits MEMS with the case $(2/3\leq \gamma\leq 1)$, hence the form of the state with $(\gamma=1)$ is more robust against amplitude damping, phase damping and depolarization channels. Here we recall that, we also observed the robustness of the MEMS with $(\gamma=1)$ with bit flip, phase flip, bit-phase-flip channels. Thus as a key result we have found that, the state MEMS remain robust against all the quantum decoherence channels with the case $(\gamma=1)$.
\section{Conclusion}
In the present paper we have investigated the influence of bit flip, phase flip, bit-phase-flip, amplitude damping, phase damping and depolarization channels on two qubits maximally entangled mixed states. The study is bifurcated in two cases related to various forms of quantum states 
with $(0\leq\gamma<2/3)$ and $(2/3\leq\gamma\leq1)$. We have found the bit flip and bit-phase-flip channels intensely effect the state and produce entanglement sudden death and birth effects in first and second cases both. On the other hand amplitude damping channel produce only entanglement sudden death with the case $(0\leq \gamma<2/3)$, but as the parameter $\gamma$ approaches to $2/3$, the amplitude damping channel keeps on reducing the entanglement sudden death zone and at last recover the entanglement in the state. The same channel has totally different behavior with the case $(2/3\leq \gamma\leq 1)$, it does not kill the entanglement in this case. As the value of the parameter $(\gamma)$ approaches to unity both the amplitude damping and phase damping channels coincides their decaying behavior of entanglement. In all the cases phase flip channel do not disturb the state. As a key result we have found, the structure of two qubits maximally entangled states with $(\gamma=1)$ is robust against all the quantum decoherence channels, it do not suffer from entanglement sudden death. We hope the present study may be useful for quantum information community.
\appendix 
\section{Operators for quantum channels}\label{aa}
\begin{center}
\begin{tabular}{|c|c|c|c|c|c|c|}
\hline 
Kraus Operators $\rightarrow$ & $E_{1}$ & $E_{2}$ & $E_{3}$ & $E_{4}$ \\ 
\hline 
Bit Flip & $\left[\begin{tabular}{c c}
$\sqrt{p}$ & $0$\\
$0$ & $\sqrt{p}$
\end{tabular}\right]$ & $\left[\begin{tabular}{c c}
$0$ & $\sqrt{(1-p)}$\\
$\sqrt{(1-p)}$ & $0$
\end{tabular}\right]$ & • & •  \\ \hline  \hline 
Phase flip & $\left[\begin{tabular}{c c}
$\sqrt{p}$ & $0$\\
$0$ & $\sqrt{p}$
\end{tabular}\right]$ & $\left[\begin{tabular}{c c}
$\sqrt{(1-p)}$ & $0$ \\
$0$ & $\sqrt{(1-p)}$ 
\end{tabular}\right]$ & • & •  \\ 
\hline \hline 
Bit Phase flip & $\left[\begin{tabular}{c c}
$\sqrt{p}$ & $0$\\
$0$ & $\sqrt{p}$
\end{tabular}\right]$ & $\left[\begin{tabular}{c c}
$0$ & $-i\sqrt{(1-p)}$\\
$i\sqrt{(1-p)}$  & $0$ 
\end{tabular}\right]$ & • & • \\ 
\hline \hline 
Amplitude damping & $\left[\begin{tabular}{c c}
$1$ & $0$\\
$0$ & $\sqrt{e^{-\gamma t}}$
\end{tabular}\right]$ & $\left[\begin{tabular}{c c}
$0$ & $\sqrt{1-e^{-\gamma t}}$\\
$0$ & $0$ 
\end{tabular}\right]$ & • & •  \\ 
\hline \hline 
Phase damping & $\left[\begin{tabular}{c c}
$\sqrt{e^{-\gamma t}}$ & $0$\\
$0$ & $\sqrt{e^{-\gamma t}}$
\end{tabular}\right]$ & $\left[\begin{tabular}{c c}
$\sqrt{1-e^{-\gamma t}}$ & $0$\\
$0$ & $0$ 
\end{tabular}\right]$ & $\left[\begin{tabular}{c c}
$0$ & $0$ \\
$0$ & $\sqrt{1-e^{-\gamma t}}$ 
\end{tabular}\right]$ & • \\ 
\hline \hline 
Depolarization & $\left[\begin{tabular}{c c}
$\sqrt{e^{-\gamma t}}$ & $0$\\
$0$ & $\sqrt{e^{-\gamma t}}$
\end{tabular}\right]$ & $\left[\begin{tabular}{c c}
$0$ & $\sqrt{\frac{1}{3}(1-e^{-\gamma t})}$ \\
$\sqrt{\frac{1}{3}(1-e^{-\gamma t})}$ & $0$ 
\end{tabular}\right]$ & $\star^{1}$ & $\star^{2}$  \\ 
\hline \hline 
\end{tabular} 
\end{center}
\begin{eqnarray}
\star^{1}=\left[\begin{tabular}{c c}
$0$ & $-i\sqrt{\frac{1}{3}(1-e^{-\gamma t})}$ \\
$i\sqrt{\frac{1}{3}(1-e^{-\gamma t})}$ & $0$
\end{tabular}\right]; \star^{2}=\left[\begin{tabular}{c c}
$\sqrt{\frac{1}{3}(1-e^{-\gamma t})}$  & $0$ \\
$0$ & $-\sqrt{\frac{1}{3}(1-e^{-\gamma t})}$
\end{tabular}\right] \nonumber
\end{eqnarray}

\end{document}